\documentclass[manuscript]{acmart}

\usepackage{algorithm}
\usepackage{algpseudocode}
\usepackage{booktabs}
\usepackage{cleveref}

\copyrightyear{2026}
\acmYear{2026}
\setcopyright{cc}
\setcctype{by}
\acmConference[HPDC '26]{The 35th International Symposium on High-Performance Parallel and Distributed Computing}{July 13--16, 2026}{Cleveland, OH, USA}
\acmBooktitle{The 35th International Symposium on High-Performance Parallel and Distributed Computing (HPDC '26), July 13--16, 2026, Cleveland, OH, USA}
\acmDOI{10.1145/3806645.3816236}
\acmISBN{979-8-4007-2640-8/2026/07}

\begin{document}

\title{Fault Tolerance of Accelerated Asynchronous Fixed-Point Iterations on Flexible Computing Infrastructure}

\author{Evan Coleman}
\affiliation{
  \institution{University of Mary Washington}
  \city{Fredericksburg}
  \state{Virginia}
  \country{USA}
}
\email{ecolema4@umw.edu}

\author{Masha Sosonkina}
\affiliation{
  \institution{Old Dominion University}
  \city{Norfolk}
  \state{Virginia}
  \country{USA}
}
\email{msosonki@odu.edu}

\renewcommand{\shortauthors}{Coleman \& Sosonkina}

\begin{abstract}
Asynchronous iterative methods tolerate straggling processors by allowing workers to proceed with stale data, but at a cost: the iterates become inconsistent, potentially degrading convergence. We investigate whether convergence accelerators such as Anderson acceleration compensate for this degradation. We experimentally study three fixed-point iterations: the Jacobi method for sparse linear systems, value iteration for the Bellman equation, and the Hartree--Fock self-consistent field (SCF) iteration. The experiments are conducted using a high-performance execution framework Ray, which abstracts the complexity of distributed systems and enables code parallelization and fault injection with minimal changes. 
 
We establish two main results. First, straggler tolerance is universal: asynchronous execution provides wall-clock speedups of $2.9\times$ (Jacobi), $7.7\times$ (VI), and $16.9\times$ (SCF) over synchronous execution with a 100\,ms-delayed worker, independent of whether acceleration is used. Second, Anderson acceleration's effectiveness under asynchrony depends on where staleness enters the computation. We identify two staleness mechanisms: iterate-level corruption, where stale worker returns directly overwrite portions of the accelerated iterate (as in block Jacobi), and evaluation-level perturbation, where staleness acts as a bounded perturbation to the fixed-point map evaluation (as in VI and SCF). Anderson acceleration fails categorically under the first mechanism but retains its benefits under the second, consistent with the perturbation analysis of Toth et al.\ (2017). This distinction, rather than the contraction norm or smoothness of the map, is the primary determinant of whether acceleration survives asynchronous execution.
\end{abstract}




\maketitle
\pagestyle{plain}

%% ===================================================================
%%  1. INTRODUCTION
%% ===================================================================
\section{Introduction}\label{sec:intro}

Fixed-point iterations, ${x}^{(k+1)} = G({x}^{(k)})$, are
found in many places throughout scientific computing. Classical iterative solvers for
linear systems (Jacobi, Gauss--Seidel, Richardson), value iteration for
Markov decision processes, and self-consistent field (SCF) iterations in
electronic structure theory are prominent examples.
In standard parallel implementations, these methods execute
in a bulk-synchronous fashion: all processors complete the current
iteration before any can begin the next. This model is straggler-sensitive (a single slow processor determines the
wall-clock time of each step), as well as fragile (any hardware fault
during an iteration typically requires discarding partial results and
restarting from a checkpoint).

As computing platforms grow more heterogeneous and adopt
cloud-based and preemptible resources, both liabilities become
increasingly costly. Asynchronous iterative methods, first analyzed
by Chazan and Miranker~\cite{chazan1969chaotic} and unified by
Frommer and Szyld~\cite{frommer2000asynchronous}, offer a possible
alternative. In the asynchronous model, each processor updates its
block of the solution using whatever information is currently available,
potentially including stale values. When the underlying map is contractive, convergence is
guaranteed under a bounded-delay assumption, although the
asynchronous iteration may converge at a different rate, or to a different fixed point.

Separately, convergence accelerators like Anderson
acceleration~\cite{anderson1965iterative} (equivalently, DIIS in
quantum chemistry~\cite{pulay1980convergence}) have long been used to
speed up synchronous fixed-point iterations. These methods maintain a
window of recent iterates and their residuals, then extrapolate to
reduce the error. Their behavior under asynchronous execution, however,
has received almost no study. This is a significant gap: if
acceleration can compensate for the inconsistencies introduced by
staleness, it could substantially broaden the regime in which
asynchronous methods are practical.

In this paper, we take a step toward closing this gap through systematic
empirical investigation. We present a Ray-based experimental
framework~\cite{moritz2018ray} for studying fault tolerance of
asynchronous fixed-point iterations under controlled injection of
delay. We study three
iterations of increasing complexity:

\begin{enumerate}
  \item \textbf{Jacobi iteration} for sparse linear systems.
        This is a linear fixed-point map whose asynchronous convergence
        is fully characterized by the spectral radius of the iteration
        matrix~\cite{chazan1969chaotic, frommer2000asynchronous}.
        The iteration contracts in the $\ell_2$-norm, and the
        Walker--Ni equivalence~\cite{walker2011anderson} gives a
        reference convergence rate for the accelerated synchronous case.
        %We use it as a controlled baseline where theoretical predictions can be compared against observed degradation under faults.

  \item \textbf{Value iteration} for the Bellman optimality equation.
        This is a nonlinear fixed-point map involving the non-smooth
        $\max$ operator, contractive in the $\ell_\infty$-norm with
        contraction constant~$\gamma$ (the discount
        factor)~\cite{bertsekas1996neurodynamic}. Asynchronous value
        iteration is well studied~\cite{bertsekas1989parallel}, but its
        combination with Anderson acceleration raises subtle theoretical
        questions: the $\ell_2$-norm least-squares problem at the heart
        of Anderson acceleration is mismatched to the
        $\ell_\infty$-contraction of the Bellman operator, and the
        non-smoothness of $\max$ violates regularity assumptions in
        existing convergence analyses~\cite{toth2015convergence}. Recent work has applied
        Anderson acceleration to synchronous value iteration~\cite{geist2018anderson, shi2019regularized,
        sun2021damped}; we look to extend this by testing it under asynchronous execution with delays.

  \item \textbf{SCF iteration} for electronic structure.
        This is a nonlinear fixed-point map whose contraction properties
        vary with problem parameters. We use the
        Pariser--Parr--Pople (PPP) model Hamiltonian~\cite{pariser1953semiempirical, pople1953electron}
        as a testbed with tunable nonlinearity. To our knowledge, no prior work studies
        SCF under asynchronous execution
\end{enumerate}

These three problems span two axes of difficulty for Anderson-type
acceleration, summarized in \Cref{tab:problem-axes}. The first
axis is the contraction norm: Jacobi and SCF contract in the
$\ell_2$-norm, where Anderson's least-squares residual minimization
is naturally well-matched, while value iteration contracts in the
$\ell_\infty$-norm, creating a norm mismatch. The second axis is the
regularity of the map: the Bellman operator's $\max$ is
non-differentiable, which invalidates the quasi-Newton interpretation
of Anderson acceleration~\cite{fang2009two} and the smoothness
assumptions in convergence proofs~\cite{toth2015convergence},
while Jacobi and SCF are both smooth. By testing the same
coordinator-level acceleration mechanism on all three, we can begin
to disentangle which theoretical properties are actually necessary
for Anderson acceleration to correct async bias.

\begin{table}[t]
  \centering
  \caption{Theoretical landscape of the three test problems.}
  \label{tab:problem-axes}
  \begin{tabular}{@{}lcccc@{}}
    \toprule
    Problem & Linear? & Contraction & Smooth? & Anderson theory \\
    \midrule
    Jacobi & Yes & $\|\cdot\|_2$ & Yes & $=$ GMRES \cite{walker2011anderson} \\
    Value iter. & No & $\|\cdot\|_\infty$ & No & Open / heuristic \\
    SCF & No & $\|\cdot\|_2$ & Yes & Local \cite{toth2015convergence} \\
    \bottomrule
  \end{tabular}
\end{table}

In all three settings, we apply Anderson acceleration at the coordinator
level; the same algorithmic mechanism under different names (Anderson
mixing for Jacobi and VI, DIIS for SCF), and study whether it can
compensate for the convergence degradation introduced by staleness. 
Our central finding is that the answer depends not on the contraction
norm or smoothness properties (\Cref{tab:problem-axes}), but on
how staleness enters the computation relative to the Anderson
subspace. We identify two distinct mechanisms:
\begin{itemize}
  \item \textbf{Iterate-level corruption} (Jacobi): stale worker
        returns directly overwrite portions of the coordinator's
        extrapolated iterate, destroying the subspace correction.
        Anderson fails at every window size and firing frequency tested.
  \item \textbf{Evaluation-level perturbation} (VI, SCF): staleness
        acts as a bounded perturbation to the fixed-point map evaluation,
        $\widetilde{G}({x}) = G({x}) + {e}$, where
        $\|{e}\|$ is controlled by the contraction constant.
        Anderson retains its acceleration properties, consistent with
        the perturbation analysis of Toth et al.~\cite{toth2017local}.
\end{itemize}
This distinction provides a practical criterion for predicting whether
acceleration will survive asynchronous execution: if the worker's
computation produces a full evaluation of the fixed-point map
(possibly on stale data), acceleration works; if it produces a
partial update that overwrites part of the iterate, acceleration
fails. Separately, the straggler tolerance of the unaccelerated
asynchronous iteration is robust across all three problems, providing
wall-clock speedups of $2.9$--$16.9\times$ at the cost of increased
total work.

%% ===================================================================
%%  2. RELATED WORK
%% ===================================================================
\section{Related Work}\label{sec:related}

% The theoretical foundations were laid by Chazan and
% Miranker~\cite{chazan1969chaotic}, who proved convergence of chaotic
% relaxation for symmetric positive definite systems.
% Baudet~\cite{baudet1978asynchronous} extended the analysis to block
% decompositions, and Bertsekas and Tsitsiklis~\cite{bertsekas1989parallel}
% provided a comprehensive treatment including asynchronous models, including establishing convergence of asynchronous value iteration.
% Frommer and Szyld~\cite{frommer2000asynchronous} unified these results
% into a common framework covering nonsingular linear systems, nonlinear
% systems, and initial value problems. More recently, Avron et
% al.~\cite{avron2015revisiting} introduced randomization, proving
% convergence rates for general SPD matrices. Chow et
% al.~\cite{chow2021asynchronous} analyzed asynchronous Richardson
% iterations, finding that optimal synchronous parameters may not remain
% optimal under asynchronous execution. Wolfson-Pou and
% Chow~\cite{wolfsonpou2019modeling} and Hook and
% Dingle~\cite{hook2018performance} developed convergence and performance
% models for asynchronous Jacobi. 
%Nayak and Anzt~\cite{nayak2024probabilistic} proposed a probabilistic framework connecting delay distributions to convergence rates.
%Spiteri~\cite{spiteri2020parallel} surveys the broader landscape.

The theoretical foundations were laid by Chazan and
Miranker~\cite{chazan1969chaotic}, extended to block decompositions
by Baudet~\cite{baudet1978asynchronous}, and unified by Frommer and
Szyld~\cite{frommer2000asynchronous}. Bertsekas and
Tsitsiklis~\cite{bertsekas1989parallel} established convergence of
asynchronous value iteration. More recent work has analyzed
convergence rates under
randomization~\cite{avron2015revisiting}, asynchronous Richardson
iterations~\cite{chow2021asynchronous}, and performance models for
asynchronous Jacobi~\cite{jensen2018using,jensen2019predictive}.

Fault tolerance for asynchronous iterative methods has been approached
from several angles. Algorithm-based fault tolerance (ABFT) techniques,
pioneered by Chen~\cite{chen2013onlineabft} for synchronous iterations
extended by Sao and Vuduc~\cite{sao2013selfstabilizing} via
self-stabilizing solvers, detect and correct soft errors through
invariant-checking and state projection. These self-stabilizing techniques have been further applied to asynchronous methods by Coleman and Sosonkina \cite{coleman2018self}. Anzt et
al.~\cite{anzt2016finegrained} showed that component-wise
update rejection transforms a synchronous Jacobi into an asynchronous
fault-tolerant iteration with modest convergence delay. Coleman et al~\cite{coleman2021fault} investigated soft-error vulnerability
of asynchronous stationary solvers directly, proposing detection and
recovery strategies based on monitoring component-wise residuals and
providing a theoretical framework for fault impact on asynchronous
convergence. These approaches target
silent data corruption (bit-flips, memory errors); our work
targets delay-induced staleness as the fault model,
motivated by cloud and heterogeneous-cluster deployments where
straggling and preemption dominate.

Anderson~\cite{anderson1965iterative} and
Pulay~\cite{pulay1980convergence} independently proposed
mathematically equivalent acceleration methods.
Walker and Ni~\cite{walker2011anderson} proved the GMRES equivalence
for linear problems; Toth and
Kelley~\cite{toth2015convergence} established local convergence of
Anderson($m$); Evans et al.~\cite{evans2020proof} proved strict rate
improvement for linearly converging methods. Fang and Saad~\cite{fang2009two} followed by 
Rohwedder and Schneider~\cite{rohwedder2011analysis} analyzed DIIS
convergence; Banerjee et al.~\cite{banerjee2016periodic} proposed
Periodic Pulay. To our knowledge, no existing electronic structure
code implements asynchronous SCF~\cite{sun2018pyscf}.
% Anderson~\cite{anderson1965iterative} proposed an acceleration method
% for fixed-point iterations based on minimizing the residual over a
% window of recent iterates. Pulay~\cite{pulay1980convergence} independently developed the mathematically equivalent
% Direct Inversion in the Iterative Subspace (DIIS) for accelerating SCF
% convergence in quantum chemistry. The equivalence was clarified by
% Fang and Saad~\cite{fang2009two}, who also connected both methods to
% quasi-Newton updating. Walker and Ni~\cite{walker2011anderson}
% showed that, for linear problems, Anderson acceleration without
% truncation is equivalent to GMRES applied to a preconditioned system, which is a
% result that implies known convergence rates. Toth and
% Kelley~\cite{toth2015convergence} proved local $r$-linear convergence
% of Anderson($m$) when the fixed-point map is contractive and the
% extrapolation coefficients remain bounded. Evans et
% al.~\cite{evans2020proof} proved that Anderson acceleration strictly
% improves the convergence rate for linearly converging methods.
% Production SCF codes~\cite{sun2018pyscf} use synchronous iteration with
% bulk-synchronous Fock builds. To our knowledge, no existing
% electronic structure code implements asynchronous SCF.
% Rohwedder and Schneider~\cite{rohwedder2011analysis} analyzed DIIS
% convergence in the synchronous setting. Banerjee et
% al.~\cite{banerjee2016periodic} proposed Periodic Pulay, applying DIIS
% every few iterations with linear mixing in between.
A result of particular relevance to our setting is that of Toth et
al.~\cite{toth2017local}, who analyzed Anderson acceleration with
inaccurate function evaluations; a model that encompasses the
stale evaluations characteristic of asynchronous execution. They showed
local improvement results persist under bounded perturbation, though
they did not study the asynchronous setting explicitly.
Zhang et al.~\cite{zhang2020globally} established
global convergence of Anderson acceleration for non-smooth
fixed-point iterations under only a nonexpansiveness assumption,
which is directly relevant to the Bellman operator's $\max$
non-smoothness.
To our knowledge, no prior work applies Anderson acceleration (or DIIS)
as a coordinator-level corrector in an asynchronous iteration.

Separately, Geist and Scherrer~\cite{geist2018anderson} first applied Anderson
acceleration to (synchronous) value iteration, reporting significant
speedups but noting the lack of theoretical guarantees. Shi et
al.~\cite{shi2019regularized} proposed Regularized Anderson Acceleration
(RAA) for off-policy deep RL. Sun et al.~\cite{sun2021damped} provided
the first convergence analysis for Anderson-accelerated policy iteration
by using the smooth MellowMax operator to replace~$\max$, showing that
smoothing restores the contraction and differentiability properties
that Anderson acceleration requires. Despite this active literature
on Anderson acceleration for synchronous RL, no prior work combines
it with asynchronous execution, which is a notable gap given that the Ray
framework underlying much modern RL was designed precisely for
asynchronous distributed computation.

%% ===================================================================
%%  3. BACKGROUND
%% ===================================================================
\section{Background}\label{sec:background}

\subsection{Asynchronous Fixed-Point Iteration}\label{sec:async-theory}

Consider a fixed-point iteration ${x}^{(k+1)} = G({x}^{(k)})$
with the solution partitioned into $p$ blocks:
${x} = ({x}_1, \ldots, x_p)$.
In the Frommer--Szyld asynchronous model~\cite{frommer2000asynchronous},
processor $l$ updates its block using a snapshot that may lag behind
the global state:
\begin{equation}\label{eq:async-model}
  {x}_l^{(k+1)} = G_l\bigl(
    {x}_1^{(s_1(k))}, \ldots, {x}_p^{(s_p(k))}
  \bigr)
\end{equation}
where $s_j(k) \leq k$ records the vintage of block~$j$ as seen by
processor $l$ at iteration~$k$. The bounded delay assumption
requires $k - s_j(k) \leq \tau$ for some finite~$\tau$.

\begin{theorem}[Frommer--Szyld~\cite{frommer2000asynchronous}]
If $G$ is a contraction on a complete metric space with contraction
constant $\rho(|{C}|) < 1$, then the asynchronous
iteration~\eqref{eq:async-model} converges to the unique fixed
point~${x}^*$ under the bounded delay assumption.
\end{theorem}

For linear systems ${A}{x} = {b}$ with the Jacobi
splitting ${A} = {D} - ({L}+{U})$, the iteration
matrix is ${M} = {D}^{-1}({L}+{U})$ and convergence
requires $\rho({M}) < 1$. For nonlinear maps such as the SCF
iteration, the relevant spectral radius is that of the Jacobian
$\partial G / \partial {x}$ evaluated at the fixed point.

\subsection{Anderson Acceleration}\label{sec:anderson}

Anderson acceleration~\cite{anderson1965iterative} maintains a window of
$m$ recent iterates $\{{x}^{(k-j)}\}_{j=0}^{m_k}$ and their
residuals ${r}^{(k)} = G({x}^{(k)}) - {x}^{(k)}$.
It computes coefficients $\{\alpha_j\}$ by solving
\begin{equation}\label{eq:anderson}
  \min_{\alpha_0,\ldots,\alpha_{m_k}}
    \Big\|\sum_{j=0}^{m_k} \alpha_j\, {r}^{(k-j)}\Big\|^2
  \quad\text{s.t.}\quad \sum_{j=0}^{m_k} \alpha_j = 1,
\end{equation}
and forms ${x}^{(k+1)} = \sum_j \alpha_j\, G({x}^{(k-j)})$.

This is identical to Pulay's DIIS~\cite{pulay1980convergence}, where
${x}^{(k)}$ is the Fock matrix~${F}_k$ and ${r}^{(k)}$
is the commutator error ${F}_k{P}_k{S} -
{S}{P}_k{F}_k$. The equivalence was established by
Fang and Saad~\cite{fang2009two} and further analyzed by
Rohwedder and Schneider~\cite{rohwedder2011analysis}.

Two results are central to our setting.
First, Walker and Ni~\cite{walker2011anderson} proved that for affine
$G({x}) = {M}{x} + {b}$, Anderson acceleration
without truncation is equivalent to GMRES applied to
$({I}-{M}){x} = {b}$. This gives convergence rate
predictions for the Jacobi case. Second, Toth et
al.~\cite{toth2017local} showed that Anderson acceleration retains
local improvement properties under inaccurate function evaluations
$\widetilde{G}({x}) = G({x}) + {e}$, provided
$\|{e}\|$ is sufficiently small, which is a model that naturally encompasses
the stale evaluations arising in asynchronous execution.

\subsection{Test Problems}
\label{sect:test-problems}
\subsubsection{Asynchronous Jacobi}
\label{sec:jacobi-prob}

For sparse ${A}{x} = {b}$ with splitting ${A} =
{D} - {L} - {U}$, the Jacobi iteration
${x}^{(k+1)} = {D}^{-1}({b} + ({L}+{U})
{x}^{(k)})$
has residual ${r}^{(k)} = {b} - {A}{x}^{(k)}$.
In our partitioned model, worker $l$ owns rows
$i \in \mathcal{B}_l$ and computes
$x_i^{(k+1)} = d_{ii}^{-1}\bigl(b_i - \sum_{j\neq i}a_{ij}
x_j^{(s_j(k))}\bigr)$.
The coordinator collects partial updates as they arrive and applies
Anderson acceleration over the global iterate/residual history.
This provides a controlled testbed since prior work helps define expected convergence rates. 
%the spectral radius $\rho({M})$ is known, the Chazan--Miranker theory predicts convergence or divergence, and the Walker--Ni equivalence~\cite{walker2011anderson} gives a reference convergence rate for the accelerated synchronous case against which the async-accelerated case can be compared.

\subsubsection{Asynchronous Value Iteration}
\label{sec:vi-prob}

A Markov decision process (MDP) is defined by states $\mathcal{S}$,
actions $\mathcal{A}$, transition probabilities $P(s'|s,a)$, rewards
$R(s,a)$, and a discount factor $\gamma \in (0,1)$. The optimal value
function $V^*$ satisfies the Bellman optimality equation
\begin{equation}\label{eq:bellman}
  V^*(s) = \max_{a \in \mathcal{A}} \Bigl[
    R(s,a) + \gamma \sum_{s' \in \mathcal{S}} P(s'|s,a)\, V^*(s')
  \Bigr].
\end{equation}
This defines a fixed-point map, $V = TV$, where the Bellman operator $T$
is a $\gamma$-contraction in the supremum
norm \cite{bertsekas1996neurodynamic}.
%$\|TV - TV'\|_\infty \leq \gamma \|V - V'\|_\infty$.
Value iteration $V^{(k+1)} = TV^{(k)}$ converges at rate~$\gamma$
for any initial~$V^{(0)}$.

In our partitioned model, the state space is divided among $p$~workers, where worker~$l$ owns states $s \in \mathcal{S}_l$ and computes
\begin{equation}
    V^{(k+1)}(s) = \max_a[R(s,a) + \gamma \sum_{s'} P(s'|s,a)\,
V^{(s_{s'}(k))}(s')],
\end{equation}
\noindent
where $s_{s'}(k)$ reflects potentially stale values of other workers'
states. Asynchronous convergence is guaranteed under bounded delay
by the $\ell_\infty$-contraction of $T$ \cite{bertsekas1989parallel}.

As noted in \Cref{tab:problem-axes}, value iteration poses two
challenges for Anderson: $\ell_2/\ell_\infty$ norm mismatch and
non-differentiability of $\max$~\cite{fang2009two, toth2015convergence}.

% Value iteration raises two theoretical challenges for Anderson
% acceleration that the other test problems do not. First, there is a
% norm mismatch: the Bellman operator contracts in
% $\|\cdot\|_\infty$ while Anderson acceleration solves the
% least-squares problem~\eqref{eq:anderson} in~$\|\cdot\|_2$. For
% Jacobi on SPD systems, the iteration matrix contracts in $\|\cdot\|_2$
% and the Walker--Ni GMRES equivalence ensures the acceleration is
% well-matched; for SCF the relevant norms are also in the $\ell_2$
% family. For value iteration, this matching fails. Second, the $\max$
% operator is non-differentiable at points where the
% optimal action switches, which invalidates the quasi-Newton
% interpretation of Anderson acceleration~\cite{fang2009two} and the
% smoothness requirements in convergence
% proofs~\cite{toth2015convergence}.

Despite these concerns, Anderson acceleration has been empirically
successful on synchronous value iteration. Geist and
Scherrer~\cite{geist2018anderson} reported significant speedups.
Sun et al.~\cite{sun2021damped} addressed the non-smoothness by
replacing $\max$ with the smooth MellowMax operator, enabling
a convergence proof for the accelerated iteration.
Zhang et al.~\cite{zhang2020globally} proved global convergence
of a stabilized Anderson variant for non-smooth nonexpansive
maps, requiring only nonexpansiveness, which is a property that the
Bellman operator satisfies (it is in fact contractive).

Our experiments test whether these empirical and theoretical results
for synchronous Anderson-accelerated VI extend to the asynchronous
setting under faults. As a sub-experiment, we also test
policy evaluation (the linear fixed-point
$V^\pi = r^\pi + \gamma P^\pi V^\pi$) which removes the $\max$
operator and gives a linear system $(I - \gamma P^\pi)V = r^\pi$.
Anderson acceleration applies cleanly via the Walker--Ni GMRES
equivalence, but the $\ell_\infty$-contraction (with constant~$\gamma$)
remains. Comparing policy evaluation against full value iteration
isolates the effect of non-smoothness from the effect of norm mismatch.

\subsubsection{Asynchronous SCF}
\label{sec:scf-prob}

The Hartree--Fock SCF iteration defines a nonlinear fixed-point map
$\mathcal{F}: {P} \mapsto {P}'$
on the one-particle density matrix~\cite{szabo1996modern},
given ${P}$, 
\begin{enumerate}
    \item builds the Fock matrix,
$F_{\mu\nu} = H_{\mu\nu} + \sum_{\lambda\sigma}P_{\lambda\sigma}
[(\mu\nu|\lambda\sigma) - \frac{1}{2}(\mu\lambda|\nu\sigma)]$,
    \item solves ${F}{C} = {S}{C}\boldsymbol{\varepsilon}$, and
    \item forms $P'_{\mu\nu} = 2\sum_{i\in\mathrm{occ}}C_{\mu i}C_{\nu i}$.
\end{enumerate}
\noindent
We partition ${P}$ by rows: worker~$\ell$ owns rows
$\mu \in \mathcal{B}_\ell$, reads the current (possibly stale)
global~${P}$, executes the full SCF map, and returns only its
owned rows of~${P}'$. The coordinator assembles, symmetrizes, and
optionally applies DIIS/Anderson acceleration.

As a model problem with full HF algebraic structure but no integral
evaluation cost, we use the Pariser--Parr--Pople (PPP)
Hamiltonian~\cite{pariser1953semiempirical, pople1953electron} for a
1D atomic chain. Each atom contributes one orthogonal basis function
(${S}={I}$), the core Hamiltonian has nearest-neighbor
hopping~$t$, and two-electron integrals use the Ohno
parameterization~\cite{ohno1964some}:
$\gamma_{\mu\nu} = U/\sqrt{1+(U R_{\mu\nu})^2}$,
where $U$ is the on-site Coulomb repulsion. The ratio $U/|t|$ controls
the correlation strength and hence the spectral radius of the SCF
Jacobian: small $U/|t|$ gives rapid contraction; large $U/|t|$ can
cause oscillation or divergence even synchronously. The PPP model is a standard benchmark for SCF convergence studies,
providing full Hartree--Fock algebraic structure (Fock build,
diagonalization, density construction) at minimal computational
cost~\cite{szabo1996modern}. It has been used to test DIIS
convergence~\cite{rohwedder2011analysis}, level-shifting strategies,
and other acceleration techniques. 
%For validation on real molecules, our framework includes a PySCF~\cite{sun2018pyscf} interface as a drop-in replacement.

\subsection{Coordinator-Level Acceleration Under Asynchrony}
\label{sec:coord-accel}

In all three test problems, the coordinator maintains a history of global
iterates and residuals and applies Anderson Acceleration via
the least-squares problem \Cref{eq:anderson}. Under asynchronous
execution, each iterate in the history is a composite: some blocks
are current while others are stale. The residual is therefore
evaluated at an inconsistent point. We identify three modes:
\begin{enumerate}
  \item Monitor-only: compute the residual norm for convergence
        monitoring but do not modify the iterates. Workers always see
        the raw asynchronously updated solution.
  \item Coordinator acceleration: extrapolate the solution at
        the coordinator and feed the accelerated iterate back to workers.
        This periodically ``resets'' the global state toward a consistent solution, limiting staleness-induced drift.
  \item Periodic acceleration: apply mode~(2) every $k$
        iterations with damped linear mixing in between, following the
        Periodic Pulay approach~\cite{banerjee2016periodic}.
\end{enumerate}
For Jacobi, mode~(2) corresponds to Anderson-accelerated Jacobi, which
is equivalent to preconditioned GMRES in the synchronous
case~\cite{walker2011anderson}. For SCF, mode~(2) is coordinator DIIS.
For value iteration, mode~(2) is Anderson-accelerated VI in the style of
Geist and Scherrer~\cite{geist2018anderson}, with the residual defined
as ${r}^{(k)} = T{V}^{(k)} - {V}^{(k)}$, the Bellman
residual.

Testing the same mechanism on all three problems, each with different
contraction norms and regularity properties, is the core experimental
design. 
%It is motivated by a structural analogy: the coordinator's
%periodic broadcast of an Anderson-extrapolated iterate is reminiscent
%of the target network mechanism in DQN~\cite{mnih2015human},
%where a frozen copy of the value function is periodically synchronized
%to stabilize training. Our framework can be seen as studying this
%synchronization mechanism in a controlled setting where ground truth is
%available.

\subsection{Staleness and Acceleration: A Coupling Perspective}
\label{sec:staleness-theory}

Consider a partitioned fixed-point iteration with $p$ workers and
coordinator-level Anderson acceleration. After the coordinator
computes the extrapolated iterate $\tilde{x}^{(k)}$, each worker
$\ell$ begins computing on a snapshot $\hat{x}^{(k)}_\ell$ that may
differ from $\tilde{x}^{(k)}$ due to communication delays. The
worker returns a partial result that the coordinator incorporates
into the global state. The question is whether the Anderson subspace
--- built from the history of these composite, partially-stale
iterates --- retains enough structure to accelerate convergence.

The key observation is that different fixed-point maps produce
partial updates with very different information content.
In value iteration, the Bellman update for a single state $s$
reads the entire value vector through the transition matrix:
$TV(s) = \max_a[R(s,a) + \gamma \sum_{s'} P(s'|s,a)\, V(s')]$.
Even though a worker returns updates for only $|\mathcal{S}_\ell|$
states, each update reflects the global state of the iterate. In
the framework of Toth et al.~\cite{toth2017local}, the staleness
acts as a bounded perturbation to the map evaluation,
$\widetilde{G}(x) = G(x) + e$, where $\|e\| \leq \rho^\tau$
is controlled by the contraction constant~$\rho$ and the staleness
depth~$\tau$. Under this bound, Anderson acceleration retains its
local improvement properties.

By contrast, in Jacobi iteration on a sparse system (e.g., a 2D
Laplacian with the 5-point stencil), the update for DOF~$i$ reads
only its immediate neighbors:
$x_i^+ = d_{ii}^{-1}(b_i - \sum_{j \in \mathcal{N}(i)} a_{ij} x_j)$.
Each worker's return encodes local boundary information, not
global structure. When these returns are assembled into the global
iterate and fed to Anderson, the subspace history is dominated by
block-boundary artifacts rather than the global error modes that
Anderson needs to build an effective extrapolation.

We refer to this property as the \textit{coupling density} of the
fixed-point map: the fraction of the full iterate that each DOF's
update depends on. Value iteration has coupling density close to~1
(every state couples to every other through~$P$). Jacobi on a
banded system has coupling density $\mathcal{O}(1/N)$ for each DOF,
though the block-level coupling (fraction of a block's
neighbors within the same block) can be higher. Our coupling
threshold experiments (\Cref{sec:jacobi-results}) show that block
internal coupling must exceed $\sim$90\% for the base iteration
to benefit from multi-sweep local solves, confirming that the
information content of each worker's return is the critical factor.

The practical implication is a design heuristic: Anderson
acceleration under asynchrony is most likely to succeed when each
worker's computation depends on a large fraction of the global
iterate (high coupling density), because the resulting partial
updates carry enough global information for the subspace to remain
well-conditioned. Coupling density thus determines which
staleness mechanism dominates: high coupling produces
evaluation-level perturbation (bounded by $\rho^\tau$), while
low coupling produces iterate-level corruption that destroys
the Anderson subspace.
When coupling is low, as in banded linear
solvers, acceleration may require alternative strategies such as
having workers return full residual evaluations rather than partial
updates, or using smaller blocks to reduce the stale fraction.

%% ===================================================================
%%  4. METHODS
%% ===================================================================
\section{Implementation}\label{sec:methods}

Our framework uses Ray~\cite{moritz2018ray} actors, one per worker,
each executing its iteration map and returning partial updates. A
coordinator actor collects results via \texttt{ray.wait()},
applies updates in arrival order, optionally runs the Anderson /
DIIS extrapolation, and relaunches workers. Fault injection is
controlled per-worker through a \texttt{FaultProfile} parameterizing
delay (mean/std), additive Gaussian noise on returned components,
drop probability, and maximum staleness. Convergence is measured
in \emph{worker-updates}, the number of partial updates applied,
which provides a consistent work metric across synchronous and
asynchronous modes.
 
A key implementation detail is the safeguard on Anderson
extrapolation. After computing the extrapolated iterate
$\tilde{x}^{(k+1)}$, the coordinator evaluates the residual and
accepts the step only if,
\begin{equation}
    \label{eq:safeguard}
    \|G(\tilde{x}^{(k+1)}) - \tilde{x}^{(k+1)}\| <
\|G(x^{(k)}) - x^{(k)}\|.
\end{equation}
\noindent
Otherwise, the un-extrapolated iterate $G(x^{(k)})$ is used.
This safeguard is essential: without it, Anderson acceleration
diverges catastrophically for value iteration (residual~$\to 10^{68}$)
due to the $\ell_2/\ell_\infty$ norm mismatch, and can produce
transient blowups for Jacobi under high staleness.
 
For Jacobi, we use the 2D Laplacian on a $100\times 100$ grid
($N=10{,}000$ unknowns) with the standard 5-point stencil,
partitioned into contiguous blocks across 4 workers.
Each worker performs 10 local Jacobi sweeps per update, which
is effective when the block's internal coupling
(fraction of a DOF's neighbors within the same block) exceeds
approximately 90\%.
For value iteration, we use Garnet random MDPs~\cite{archibald1995generation}
with $|\mathcal{S}| \in \{200, 500, 1000\}$,
$|\mathcal{A}| \in \{4, 10\}$, branching factor~$b=5$, and
$\gamma \in \{0.9, 0.95, 0.99\}$. Grid-world navigation problems
with known optimal policies serve as validation.
For SCF, we use the PPP chain with $n_\text{atoms} = 8$ at
$U/|t| \in \{2, 2.5\}$ and $n_\text{atoms} = 20$ at $U/|t| = 4$.
Straggler experiments inject per-worker delays of 5--100\,ms on
one or two workers; all experiments are repeated over multiple
fault realizations. All experiments were run on the ACES HPC cluster at
Texas A\&M University using 2 nodes with 8 CPUs each.

%% ===================================================================
%%  5. RESULTS
%% ===================================================================
\section{Results}\label{sec:results}
 
\subsection{Jacobi Iteration}\label{sec:jacobi-results}

\begin{figure}[t]
  \centering
  \includegraphics[width=\linewidth]{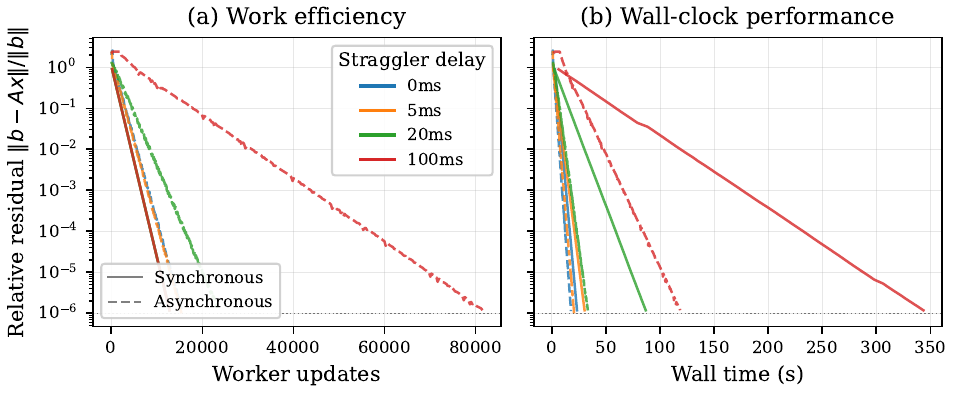}
  \caption{Jacobi straggler tolerance. Left: residual vs.\ worker-updates
  (sync curves overlap; async fans right). Right: residual vs.\ wall time
  (sync fans right; async stays compact). Solid = synchronous;
  dashed = asynchronous.}
  \label{fig:straggler}
\end{figure}

\Cref{tab:straggler} summarizes the straggler tolerance
experiments on a $100\times100$ Laplacian with 4 workers, where one
worker is artificially delayed. Synchronous iteration always converges
in exactly 12,960 worker-updates (WU),  equal 3,240 rounds $\times$ 4 workers, 
regardless of delay; the algorithm is deterministic and delay only
affects wall time. Asynchronous iteration needs progressively more
work (up to $6.3\times$) because the straggler's DOFs become stale,
but the wall-clock savings dominate: $2.9\times$ faster even at 100\,ms
delay (\Cref{fig:straggler}).
 
\begin{table}[t]
  \centering
  \caption{Straggler tolerance: sync vs.\ async Jacobi (100$\times$100 grid, 4 workers). All experiments converge.}
  \label{tab:straggler}
  \begin{tabular}{@{}lrrrrr@{}}
    \toprule
    Delay & \multicolumn{2}{c}{Synchronous} & \multicolumn{2}{c}{Asynchronous} & Wall \\
    \cmidrule(lr){2-3}\cmidrule(lr){4-5}
    & WU & Time & WU & Time & speedup \\
    \midrule
    0\,ms  & 12{,}960 & 23.4\,s  & 15{,}610 & 17.6\,s  & 1.3$\times$ \\
    5\,ms  & 12{,}960 & 30.5\,s  & 15{,}400 & 20.4\,s  & 1.5$\times$ \\
    20\,ms & 12{,}960 & 87.8\,s  & 23{,}710 & 33.2\,s  & 2.6$\times$ \\
    100\,ms& 12{,}960 & 348.0\,s & 81{,}535 & 118.6\,s & 2.9$\times$ \\
    \bottomrule
  \end{tabular}
\end{table}
 
\begin{figure}[t]
  \centering
  \includegraphics[width=\linewidth]{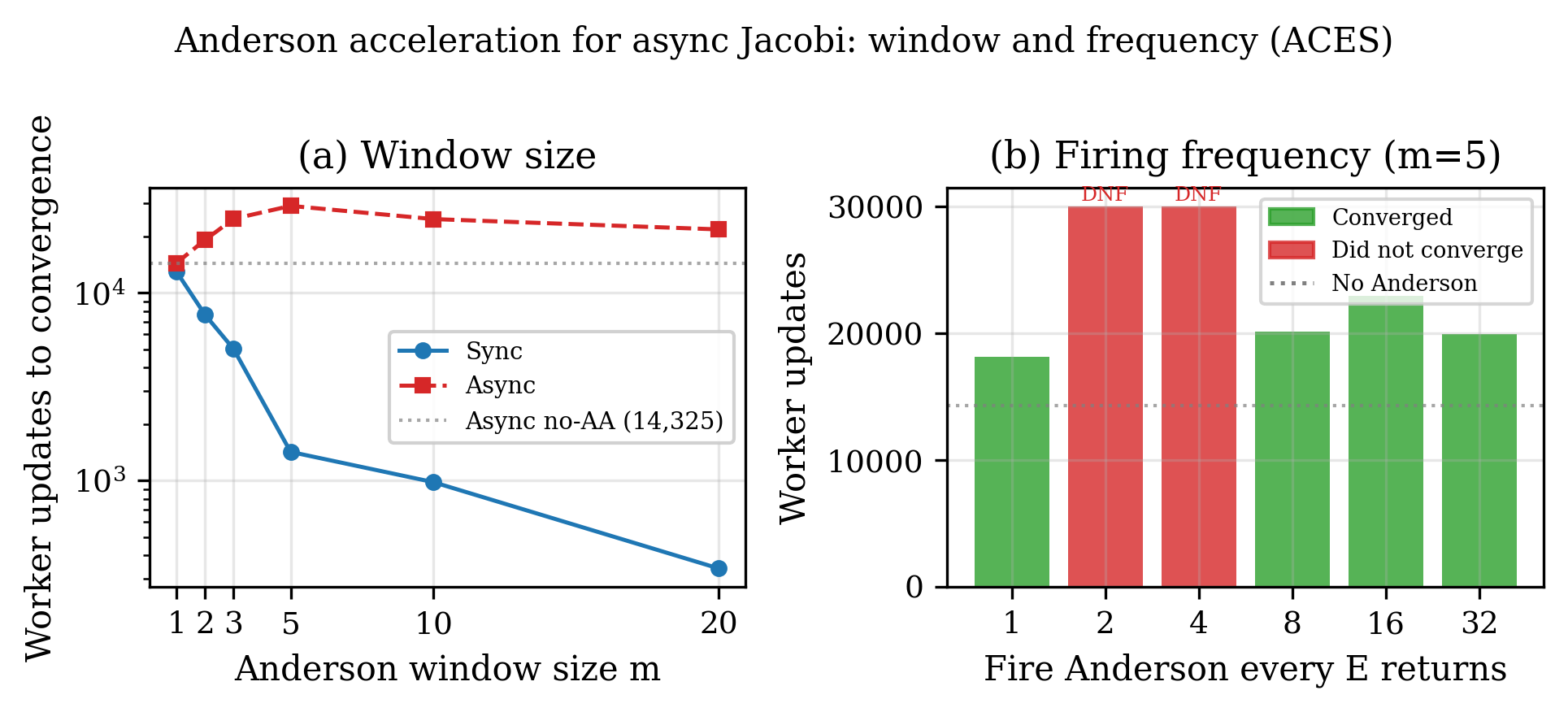}
  \caption{Anderson acceleration for async Jacobi.
  (a)~Window size: sync improves monotonically; async always above the
  no-Anderson baseline (gray line).
  (b)~Firing frequency: $E=2,4$ diverge; $E \geq 8$ converge but do
  not beat no-Anderson.}
  \label{fig:anderson_sweep}
\end{figure}

In synchronous mode, Anderson acceleration is spectacularly effective:
Anderson(20) reduces the iteration count by $38\times$ (85 vs.\ 3,240
iterations), consistent with the Walker--Ni GMRES
equivalence~\cite{walker2011anderson}. In asynchronous mode, however,
Anderson actively hurts convergence regardless of tuning
(\Cref{fig:anderson_sweep}). We swept both the window size
$m \in \{1,2,3,5,10,20\}$ and the firing frequency (every $E \in
\{1,2,4,8,16,32\}$ worker returns). Every combination requires more
worker-updates than plain async Jacobi (\Cref{tab:straggler}). Firing every
$E=2$ or $E=4$ returns causes divergence; firing less frequently
($E \geq 8$) converges but still requires $40$--$60\%$ more work than
the un-accelerated baseline. The mechanism is clear: when Anderson
extrapolates the global iterate, workers with in-flight updates
computed on the pre-extrapolation state return stale results
that partially overwrite the extrapolated solution.
 
This is consistent with the coupling density analysis of
\Cref{sec:staleness-theory}: each Jacobi worker's return
encodes only local boundary information (low coupling), so
the Anderson subspace history is dominated by block-boundary
artifacts rather than global error modes. No amount of tuning
recovers the Anderson-acceleration benefit because the information deficit is structural.
 
\begin{figure}[t]
  \centering
  \includegraphics[width=0.55\linewidth]{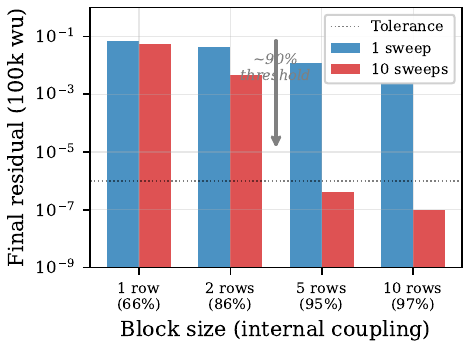}
  \caption{Block coupling threshold. Multi-sweep local solves become
  transformative above $\sim$90\% internal coupling.}
  \label{fig:coupling}
\end{figure}

We identify a sharp threshold for multi-sweep local solves
(\Cref{fig:coupling}): block internal coupling must exceed
$\sim$90\% for multiple sweeps to be effective. Below this threshold
(1--2 rows, 66--86\% coupling), 10 sweeps provide negligible benefit;
above it (5--10 rows, 95--97\%), they yield a $30{,}000\times$
improvement, explaining the block design choice
in~\cite{chow2015asynchronous}.
 
\subsection{Value Iteration}\label{sec:vi-results}
 
We test on Garnet MDPs with $|\mathcal{S}| \in \{200, 500, 1000\}$
and $\gamma \in \{0.9, 0.95, 0.99\}$, using 4 workers and 10
random fault realizations per configuration.

\begin{figure}[t]
  \centering
  \includegraphics[width=\linewidth]{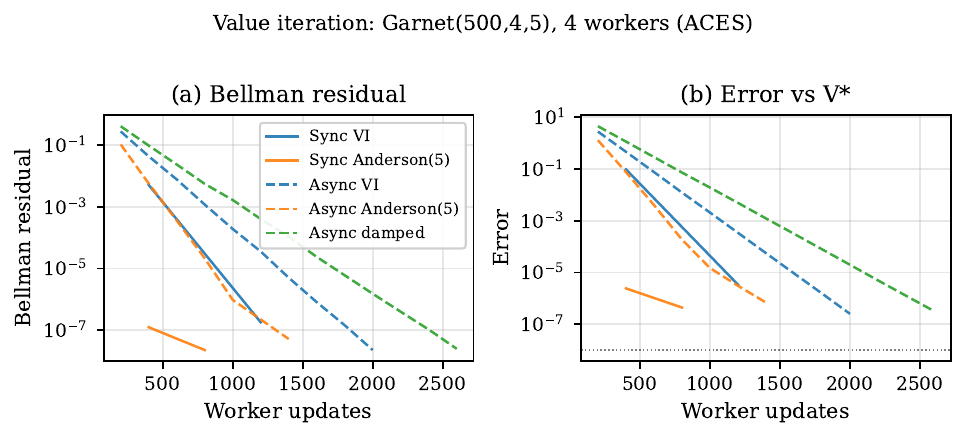}
  \caption{VI convergence on Garnet(500,4,5,$\gamma$=0.95).
  Anderson(5) accelerates both sync and async. Damping hurts.
  Dashed = async (median of 10 realizations).}
  \label{fig:vi_convergence}
\end{figure}

\begin{figure}[t]
  \centering
  \includegraphics[width=\linewidth]{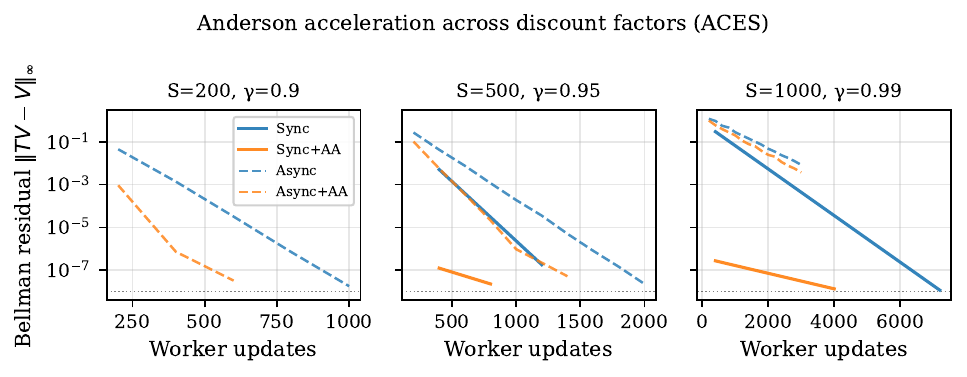}
  \caption{Anderson benefit grows with $\gamma$. At $\gamma=0.99$,
  sync Anderson reduces iterations by $1.7\times$ but async needs
  more budget. Orange = Anderson; blue = plain.}
  \label{fig:vi_gamma}
\end{figure}

Unlike the Jacobi case, safeguarded Anderson(5) provides a consistent
$1.2$--$1.7\times$ iteration reduction for value iteration in
both synchronous and asynchronous modes
(\Cref{fig:vi_convergence}). The benefit grows with the
discount factor: $1.2\times$ at $\gamma=0.9$, $1.4\times$ at
$\gamma=0.95$, and $1.7\times$ at $\gamma=0.99$ in synchronous mode
(\Cref{fig:vi_gamma}). In asynchronous mode, Anderson provides
a $\sim$$1.3\times$ speedup at $\gamma \leq 0.95$; at $\gamma=0.99$
the perturbation bound loosens and Anderson still helps but requires
a larger iteration budget (Figure~\ref{fig:vi_gamma}).
Damping ($\alpha=0.3$) uniformly hurts convergence compared to plain
async VI.
 
This is consistent with the coupling density analysis of
\Cref{sec:staleness-theory}: the Bellman operator couples each
state to the full value vector through the transition matrix,
so each worker's partial update carries global information.
Staleness enters as a bounded perturbation
$\|e\| \leq \gamma^\tau$ to the map evaluation, within the
regime analyzed by Toth et al.~\cite{toth2017local}. The bound
tightens with~$\gamma$, explaining why Anderson's benefit grows
with the discount factor while simultaneously predicting that
it will diminish as $\gamma \to 1$.
 
\begin{figure}[t]
  \centering
  \includegraphics[width=\linewidth]{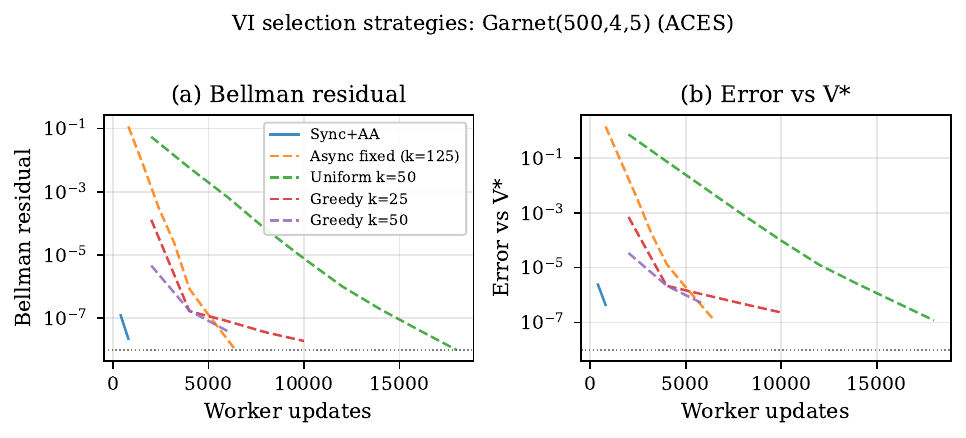}
  \caption{VI selection strategies. Greedy outperforms uniform at
  every~$k$. Fixed partition is fastest in total work.}
  \label{fig:vi_selection}
\end{figure}

Selection strategy interacts with coupling density
(\Cref{sec:staleness-theory}). For Jacobi, block size is
the primary factor and greedy selection does not help, likely because
residuals are dominated by block-boundary effects. For VI, greedy
selection converges in 2,680 iterations vs.\ $>$5,000 for uniform
at $k=25$, suggesting that which states to update matters more
than how many; consistent with VI's high coupling density propagating
corrections globally. Fixed-partition async ($k=125$) converges
fastest in total work (1,680 iterations) by avoiding selection
overhead.
 
\begin{figure}[t]
  \centering
  \includegraphics[width=\linewidth]{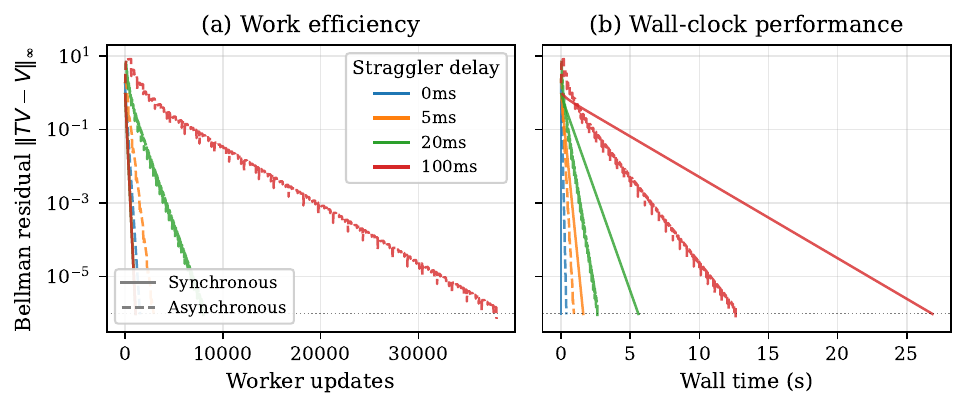}
  \caption{VI straggler tolerance on Garnet(500,4,5,$\gamma$=0.95),
  ACES data. Same pattern as Jacobi: sync curves fan out on wall
  time (right); async stays compact.}
  \label{fig:vi_straggler}
\end{figure}
 
\subsection{SCF Iteration}
\label{sec:scf-results}
 
\begin{figure}[t]
  \centering
  \includegraphics[width=\linewidth]{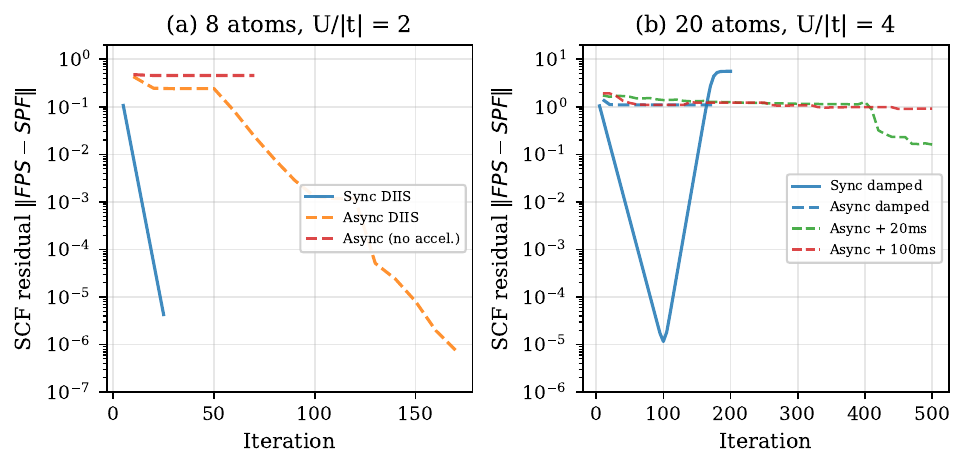}
  \caption{SCF convergence. (a)~8 atoms, $U/|t|=2$: DIIS corrects
  async bias (177 vs.\ 28 iterations). (b)~20 atoms, $U/|t|=4$:
  damped SCF shows wall-clock straggler tolerance despite
  incomplete convergence.}
  \label{fig:scf_combined}
\end{figure}

We test the PPP Hamiltonian at two regimes of correlation strength
(\Cref{fig:scf_combined}).

Synchronous SCF+DIIS converges in 28 iterations to machine
precision ($\Delta E = 2.8 \times 10^{-14}$~eV). Without DIIS,
asynchronous SCF converges to a different fixed point with
$0.6$\,eV energy error. With coordinator-level DIIS, however,
asynchronous SCF converges to the \emph{correct} energy with error
$6.5 \times 10^{-7}$\,eV in 177 iterations---$6.3\times$ more
iterations than sync, but the same correct answer. This demonstrates
that DIIS can correct the async bias for SCF at weak correlation,
at the cost of increased iteration count.

\begin{figure}[t]
  \centering
  \includegraphics[width=\linewidth]{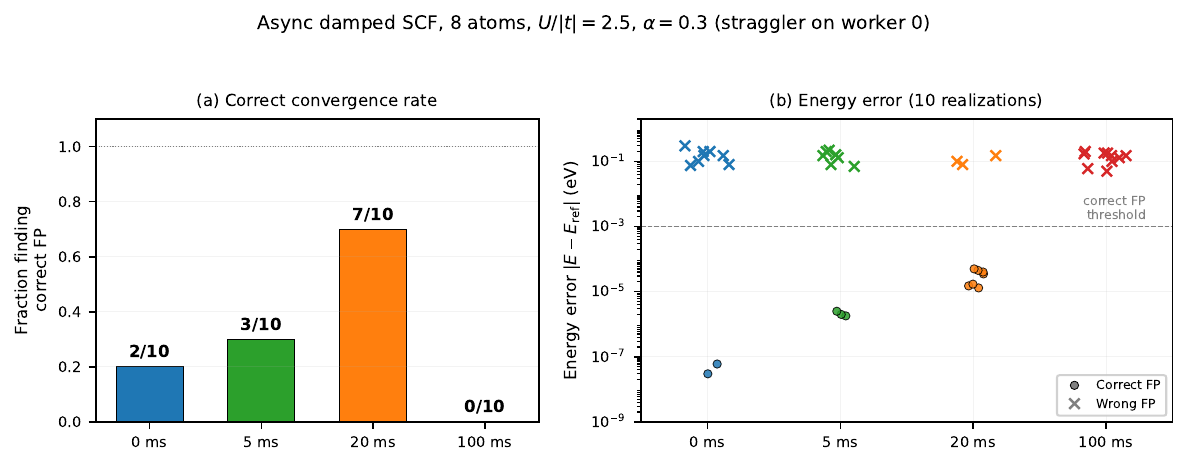}
  \caption{Stochastic SCF convergence at $U/|t|=2.5$ with damped
  mixing (10 realizations per delay). (a)~Fraction finding the
  correct fixed point. (b)~Energy error per realization; circles =
  correct FP, crosses = wrong FP.}
  \label{fig:scf_stochastic}
\end{figure}

At intermediate correlation ($U/|t|=2.5$, 8 atoms), the SCF energy
landscape admits multiple fixed points and async convergence becomes
stochastic: different Ray scheduling realizations lead to different
solutions. We ran 10 realizations of damped async SCF ($\alpha=0.3$)
at each straggler delay (Figure~\ref{fig:scf_stochastic}).
With no straggler, only 2/10 realizations find the correct fixed
point; the other 8 converge to 6 distinct spurious solutions with
$\Delta E$ ranging from $0.05$ to $0.48$\,eV. Counterintuitively,
moderate delay \emph{improves} reliability: at 20\,ms, 7/10
realizations find the correct energy ($\Delta E \sim 10^{-5}$\,eV).
The likely mechanism is that the straggler effectively serializes
one worker's contributions, reducing the inconsistency among the
remaining workers and narrowing the basin of attraction toward the
correct solution.
At 100\,ms delay, no realization finds the correct fixed point:
the perturbation bound is exceeded regardless of scheduling.
 
At strong correlation ($U/|t|=4$, 20 atoms), even synchronous DIIS
diverges and damped SCF barely converges, but the straggler tolerance
remains dramatic: $16.9\times$ wall-clock speedup at 100\,ms delay.
 
All three regimes are consistent with the coupling density
framework of \S\ref{sec:staleness-theory}: each worker computes
the full SCF map on a stale density snapshot, coupling each
basis function to all others through the two-electron integrals.
The perturbation bound $\|e\| \leq \rho^\tau$ depends on the
spectral radius~$\rho$ of the SCF Jacobian.
At $U/|t|=2$, $\rho$ is small and DIIS retains its corrective
power. At $U/|t|=2.5$, $\rho$ is large enough that the iteration
is near the basin boundary---scheduling noise determines which
fixed point is reached---explaining both the stochasticity and
the non-monotonic effect of delay.
 
\subsection{Analysis}
\label{sec:analysis}
 
\Cref{tab:cross} summarizes the experimental results through
the lens of coupling density (\Cref{sec:staleness-theory}).
Straggler tolerance (row~1) is universal and independent of the
acceleration question, depending only on the throughput advantage
of asynchronous execution.

The acceleration results (rows~2--4) are explained by the information
content of each worker's return. In VI and SCF, high coupling means
each partial update reflects the full iterate; staleness acts as a
bounded perturbation to the map evaluation, and Anderson/DIIS retains
its benefits when the contraction is sufficiently strong
($\gamma \leq 0.95$ for VI, small $U/|t|$ for SCF). The
$U/|t|=2.5$ SCF experiments further show that near the basin
boundary, convergence becomes stochastic: scheduling order
determines which fixed point is reached, and moderate delay
can paradoxically improve reliability by partially serializing
worker contributions. In Jacobi on
a sparse system, each worker's return encodes only local boundary
information, and Anderson's subspace cannot capture the global error
modes needed for effective extrapolation; regardless of window size
or firing frequency.

This resolves an apparent puzzle: why does Anderson fail for
Jacobi (a linear, $\ell_2$-contractive, smooth problem with
well-understood theory) but succeed for VI (a nonlinear,
$\ell_\infty$-contractive, non-smooth problem with weaker
guarantees)? The classical properties in
\Cref{tab:problem-axes} govern the synchronous
iteration. Under asynchrony, the relevant property is how much
global information each worker's computation carries---a property
of the problem's  structure, not its contraction
norm or smoothness. An open question is whether low-coupling
problems like Jacobi can be restructured (e.g., having workers
return full residual vectors rather than solution updates, or
using overlapping blocks) to recover acceleration benefits.

\begin{table}[t]
  \centering
  \caption{Cross-problem comparison through the coupling density lens.}
  \label{tab:cross}
  \begin{tabular}{@{}lccc@{}}
    \toprule
    & Jacobi & Value iter. & SCF \\
    \midrule
    Straggler speedup  & $2.9\times$ & $7.7\times$ & $16.9\times$ \\
    Coupling density    & Low & High & High \\
    Anderson sync      & $38\times$ & $1.2$--$1.7\times$ & DIIS: 28 iter \\
    Anderson async     & Hurts & $1.3\times$ & Depends on $\rho, \tau$ \\
    \bottomrule
  \end{tabular}
\end{table}

%% ===================================================================
%%  6. CONCLUSION
%% ===================================================================
\section{Conclusion}\label{sec:conclusion}
 
The critical factor is coupling density: how much of the
global iterate does each worker's computation depend on?
High coupling (VI, SCF) produces evaluation-level perturbation
bounded by~$\rho^\tau$, and Anderson/DIIS retains its acceleration
within the regime predicted by Toth et al.~\cite{toth2017local}.
Low coupling (Jacobi) produces iterate-level corruption, and no
tuning of window size or firing frequency recovers the benefit.
This explains why Anderson fails for the mathematically ``easiest''
problem (linear, smooth Jacobi) but succeeds for the ``hardest''
(nonlinear, non-smooth VI): the relevant property is coupling
structure, not contraction norm or smoothness.
 
This distinction is a property of the system design---how workers
partition and return their work---not the underlying mathematics,
providing a practical criterion for predicting whether acceleration
will survive asynchronous deployment.
Separately, we confirm that the straggler tolerance of
unaccelerated asynchronous iteration is universal across all three
problems, providing wall-clock speedups of $2.9$--$16.9\times$ at the
cost of $1.2$--$6.3\times$ more total work. This tradeoff is
independent of the acceleration question and represents the primary
practical value of asynchronous methods on flexible infrastructure.
Two additional findings provide practical design guidance.
First, for spatially-coupled problems, multi-sweep local solves
require block internal coupling above a sharp $\sim$90\% threshold to
be effective. Second, greedy (Gauss--Southwell) state selection
substantially outperforms uniform random selection for value iteration.
 
Future work will extend this framework to larger-scale cloud
deployments with naturally occurring straggler distributions, and
will investigate whether iterate-corruption problems can be redesigned
(e.g., by having workers return full fixed-point map evaluations
rather than partial updates) to recover acceleration benefits.

%==============================================================================
% Acknowledgments
%==============================================================================
\section*{Acknowledgments}
This work leveraged the ACES Cluster at Texas A\&M University under allocation CIS250436 from the Advanced Cyberinfrastructure Coordination Ecosystem: Services \& Support (ACCESS) program, which is supported by U.S. National Science Foundation grants \#2138259, \#2138286, \#2138307, \#2137603, and \#2138296.

%% ===================================================================
%%  REFERENCES
%% ===================================================================
\bibliographystyle{ACM-Reference-Format}
\bibliography{references}

\end{document}